\documentclass[10pt]{iopart}
\usepackage{graphicx,amssymb}
\begin{document}

\title{Calculated vibrational and electronic properties of various sodium thiogermanate glasses}
\author{S\'ebastien Blaineau and Philippe Jund }

\address{Laboratoire de Physicochimie de la Mati\`ere Condens\'ee, Universit\'e Montpellier 2, Place E. Bataillon, Case 03, 34095 Montpellier, France}

\begin{abstract}
We study the vibrational and electronic properties of (x)Na$_2$S-(1-x)GeS$_2$ glasses through DFT-based molecular dynamics simulations, at different sodium concentrations ($0<x<0.5$). We compute the vibrational density of states for the different samples in order to determine the contribution of the Na$^+$ ions in the VDOS. With an in-depth analysis of the eigenvectors we determine the spatial and atomic localization of the different modes, and in particular in the zone corresponding to the Boson peak. We also calculate the electronic density of states as well as the partial EDOS, in order to determine the impact of the introduction of the sodium modifiers on the electronic properties of the GeS$_2$ matrix.
\end{abstract}
PACS numbers: 61.43.Fs,61.43.Bn,71.23.Cq,71.23.-k,71.15.Pd
 

\section{Introduction}
Sodium thiogermanate (x)Na$_2$S-(1-x)GeS$_2$ glasses are good solid electrolytes, with a high ionic conductivity at room temperature \cite{robinel,tranchant}.  As in other glassy systems, the ionic transport process has been clearly determined but its microscopic origin is still not well understood. In
particular the mechanisms leading to the high conductivity at room temperature are not clearly
established: is the conductivity dominated by random back and forth jumps \cite{funke} or do preferential
pathways (``channels'') exist inside the glassy matrix as in amorphous silica \cite{jund} ? A first
step in the comprehension of these mechanisms is to determine the effect of the modifier ions (Na)
on the physical properties of the glass (GeS$_2$). To that purpose, theoretical studies, and in particular molecular dynamics (MD) simulations are interesting tools that provide detailed informations at the atomic level on  the modifications of the amorphous sample as the concentration of sodium is 
increased. These modifications have then to be connected to the evolution of the ionic transport if
this connection exists.\\
In previous works we have studied germanium disulfide glasses (GeS$_2$) through DFT-based MD simulations \cite{blaineau1,blaineau2,blaineau3}. The results provided by our simulations were in very good agreement with the existing experimental data. Furthermore additional informations at the atomic scale 
provided by these simulations were found to explain some properties observed experimentally at the macroscopic scale especially concerning the vibrational and electronic properties of GeS$_2$ 
glasses.\\
In this paper we aim to analyze Na-Ge-S systems using the same model, in order to study the impact of the Na$^+$ ions in GeS$_2$ glasses. Although similar studies have been performed in other glassy systems, such as SiO$_2$ \cite{jund}, showing the evidence of conduction channels created dynamically 
by the Na cations inside the glass, no MD simulations have been performed yet in sodium thiogermanate glasses to our knowledge. The aim is {\em in fine} to analyze the influence of the glassy matrix on
the conduction properties: do the conduction channels exist also in chalcogenide glasses or are
they specific to oxide glasses ? Are the modifications introduced by the alkali ions similar in 
both types of glasses ?  To that purpose, we focus here (in a first step) on the vibrational and 
electronic properties of Na-Ge-S glasses, which should be directly connected to the ionic conduction properties. In order to evaluate the evolution of theses properties with the concentration of alkali ions, we simulate several (x)Na$_2$S-(1-x)GeS$_2$ samples, for $0<x<0.5$, and analyze the vibrational and electronic densities of states for these different Na concentrations.
The article is organized as follows : In section II we briefly describe the theoretical foundations of our model, whereas in section III we study the vibrational properties of the amorphous samples through an in-depth analysis of the vibrational eigenvectors. Subsequently we study the electronic properties in section IV, and finally in section V we summarize the major conclusions of our work. 
\section{Model}

The code we have used is a first-principles type molecular dynamics program 
called FIREBALL96, which is based on the local orbital electronic structure 
method developed by Sankey and Niklewski \cite{sankey}. The foundations of 
this model are the Density Functional Theory (DFT) \cite{dft} within the 
Local Density Approximation (LDA) \cite{lda}, and the non-local 
pseudopotential scheme \cite{pseudo}. The use of the non-self-consistent 
Harris functional \cite{harris}, with a set of four atomic orbitals 
(1$s$ and 3$p$) per site that vanish outside a cut-off radius of $5a_0$ 
(2.645~\AA ) considerably reduces the CPU time.\\
The pseudo-wave function $\Psi$ of the system is given by the following equation:
\begin{eqnarray}
\Psi_j(\vec{k},\vec{r})=\Sigma_{\mu}C_{\mu}^j(\vec{k})\Phi_{fireball}^{\mu}(\vec{r})
\end{eqnarray}
where j is the band index, $\Phi_{fireball}^{\mu}$ is the fireball basis function for orbital $\mu$, and C$_{\mu}^j(\vec{k})$ are the LCAO expansion coefficients. Only the $\Gamma$ point is used to sample the
Brillouin zone ($\vec{k}=\vec{0}$).\\
This model has given excellent results in many different 
chalcogenide systems over the last ten years \cite{blaineau1,drabold,junli}.
In the present work we melt a crystalline $\alpha$-GeS2 configuration containing 258 particles 
at 2000K during 60 ps (24000 timesteps) in a cubic box of 19.21 \AA, until we obtain an 
equilibrated liquid.
Subsequently we replace randomly GeS$_4$ tetrahedral units by artificial Na$_2$S$_3$ ``molecules'', following a procedure similar to the one used in SiO$_2$ glasses \cite{jund}, in order to obtain a given sodium concentration (the total number of atoms, N, is kept constant at 258). We generate thus eight (x)Na$_2$S-(1-x)GeS$_2$ samples at different sodium concentrations (x= 0, 0.015, 0.03, 0.06, 0.11, 0.2, 0.33 and 0.5).
 The bounding box is rescaled each time so that the density matches its experimental counterpart (from 19.21 \AA~ for  x=0\cite{boolchand} to 18.3 \AA~ for x=0.5\cite{ribes}), in order to limit artificial pressure effects on the system. Then, we melt the resulting system at 2000K during 60 ps so that the system completely loses the memory of the initial artificial configuration, and becomes a homogeneous liquid (x)Na$_2$S-(1-x)GeS$_2$ system.
Finally, we quench the liquid structure at a quenching rate of 6.8$\times$10$^{14}$~K/s, decreasing the temperature to 300K through the glass transition temperature T$_g$, and we let our sample relax at 300K during 100 ps. The dynamical matrix and the electronic density of states have been calculated at the end of this relaxation time. It is worth noticing that the results obtained with 
FIREBALL96 on several Na-Ge-S test samples are almost identical to those obtained with the self-consistent {\em ab initio} 
SIESTA code \cite{siesta} in which the largest available basis set has been used. This shows the ability of our model to 
accurately describe sodium thiogermanate glasses at a relatively reduced CPU time cost.
\section{Results}

\subsection{Vibrational properties}

First we compute the Vibrational Density of States (VDOS), which can be measured experimentally by inelastic neutron diffraction spectroscopy. The VDOS is calculated through the diagonalization of D, the dynamical matrix of the system given by:

\begin{eqnarray}
D(\phi_i,\phi_j)=\frac{\partial^2 E(\phi_i,\phi_j)}{\partial\phi_i\partial\phi_j} , ~~\phi=x,y,z
\end{eqnarray}
for two particles $i$ and $j$. Fig.1 presents the calculated VDOS for different values of x  ( for clarity only the results for  x=0, x=0.2 and x=0.5 are shown).
The VDOS of GeS$_2$ (x=0) has been studied in detail in a previous work \cite{blaineau2} in which we have determined the presence of two bands (the acoustic and optic band), separated by a "gap", which was found to contain a few localized modes caused by bond defects.
We can see in Fig. 1 how the introduction of sodium atoms modifies the VDOS, and it appears clearly that the vibrational contribution of the Na atoms takes place between the acoustic and optic band (200 cm$^{-1}$- 300 cm$^{-1}$). In the Na$_2$GeS$_3$ compound (x=0.5) the two bands cannot be distinguished anymore, since the density of states is almost flat over the whole spectrum. It can also be seen that the introduction of sodium cations diminishes the acoustic band especially on the low frequency side, contrarily to the optic band that remains practically unchanged. The diminution of the low frequency modes is counterbalanced by an accumulation of modes in the ``gap'' zone between 200 cm$^{-1}$ and 300 cm$^{-1}$ when the sodium concentration increases.
The low-frequency zone at 35 cm$^{-1}$ has been attributed experimentally to the well-known ``Boson peak'', which is a signature of amorphous materials in the VDOS \cite{philips}.  It appears in our simulation that above x=0.2, the density of states in that zone decreases significantly (together with the rest of the acoustic band). Unfortunately, the  VDOS of sodium thiogermanate glasses has never been measured experimentally to our knowledge, and therefore we cannot confirm this lack of low frequency modes by experimental data.

In order to measure the localization of the vibrational modes, we calculate the participation ratio P$_r$ \cite{bell2}:
\begin{eqnarray}
P_r=\frac{(\Sigma_{i=1}^N|\vec{e}_i(\omega)|^2 )^2}{N\Sigma_{i=1}^N|\vec{e}_i(\omega)|^4}
\end{eqnarray}
where the summation is done over the N particles of the sample. If the mode corresponding to 
eigenvalue $\omega$ is delocalized and all atoms vibrate with equal amplitudes, then P$_r$($\omega$) will be close to 1. On the contrary, if the mode is strongly localized, then 
P$_r$($\omega$) will be close to 0. 
The results are shown in Fig.2, and it can be seen that the modes that appear in the zone between 200 cm$^{-1}$ and 300 cm$^{-1}$ in the sodium-enriched samples are relatively delocalized. The P$_r$, which is close to zero for GeS$_2$ in that region, becomes higher  as the Na concentration increases.
 An in-depth study of the vibrational eigenvectors in that region shows that these modes are mainly caused by sodium atoms, as it could be deduced from the VDOS (Fig.1). Although a few localized modes appear at the end of the optic band (beyond 480 cm$^{-1}$), it can be said that the contribution of the Na atoms in the VDOS is principally sensitive in delocalized modes (for x=0.5, the maximum of Pr appears at 190 cm$^{-1}$).
It should also be noted that in the region attributed to the Boson Peak (35 cm$^{-1}$) \cite{tanaka}, the participation ratio is lower for sodium-enriched systems. Since we have previously seen that this zone showed a lack of modes for x=0.33 and x=0.5 in comparison to the low values of x one can conclude that the remaining "soft modes" become more localized when the Na  concentration increases.

In order to calculate the spatial localization of the vibrational modes, we calculate the center of gravity $\vec{r}_g(\omega)$ of each mode of eigenvalue $\omega$, and the corresponding ``localization'' length L \cite{zotov}, as
\begin{eqnarray}
\vec{r}_g(\omega)=\frac{\Sigma_{i=1}^N\vec{r}_i|\vec{e}_i(\omega)|^2/m_i}{\Sigma_{i=1}^N |\vec{e}_i(\omega)|^2/m_i}
\end{eqnarray}
and
\begin{eqnarray}
L(\omega)=\sqrt{\frac{\Sigma_{i=1}^N|\vec{r}_i-\vec{r}_g(\omega)|^2|\vec{e}_i(\omega)|^2/m_i}{\Sigma_{i=1}^N|\vec{e}_i(\omega)|^2/m_i}}
\end{eqnarray}
where $\vec{r}_i$ and $m_i$ are respectively the position and the atomic mass of particle $i$. Periodic boundary conditions must be taken into account in these calculations. The localization length (Fig.3) represents the spatial localization of a given mode, and its maximal value corresponds to the half-size of the box. Beyond this length, the amplitude of the atomic vibrations decreases significantly. 
It can be seen in Fig.3 that contrarily to the participation ratio, the localization length remains unchanged at low frequencies as the Na concentration increases, and corresponds approximately to the half-size of the box. The low frequency modes appear therefore completely delocalized in space, but the 
number of particles involved in these vibrations decreases as the sodium concentration becomes 
higher as shown by the decreasing participation ratio. This means that in pure GeS$_2$ a large
amount of connected particles were involved in the low frequency modes whereas in Na-Ge-S glasses, Na
breaks these connections and therefore small groups of particles (small participation ratio) 
scattered in the whole simulation box (large localization length) participate in these modes with 
the consequence of a decrease of the density of states at low frequency.

\subsection{Electronic properties}

	The Sankey-Niklewski scheme that has been described in section II allows the
	determination of the electronic energy eigenvalues for the different samples.
	The Electronic Density of States (EDOS), which can be calculated by ``binning'' these
	eigenvalues, has been measured experimentally by X-ray
	Photoelectron Spectroscopy (XPS) in GeS$_2$ and Na$_2$GeS$_3$ (x=0.5) systems \cite{foix}.
In a previous work we have compared in GeS$_2$ our calculated EDOS with its experimental counterpart, and we have analyzed in detail the different features of the valence spectrum \cite{blaineau3}. Three bands, called A, B and C were clearly distinguished, with good agreement with the XPS spectrum.
 We show in Fig.4 the valence band of our calculated EDOS for x=0 and x=0.5, whereas Fig.5 presents the calculated and experimental valence spectra of Na$_2$GeS$_3$. 
We can see in Fig.4. that the impact of the sodium atoms in the EDOS of amorphous GeS$_2$ is negligible (the spectra of the other simulated samples are rather similar to these two graphs). It can however be seen that the width of band B decreases with the introduction of the sodium atoms in the glassy sample and that the density of states in band C increases. In addition, the small peak at the end of band A is slightly shifted in the sodium-enriched sample. We can see in Fig.5 that this calculated spectrum is in good agreement with the experimental data, even though a small energy shift is visible at the end of band C ($\approx$ 1 eV).\\
In order to determine which atomic orbitals are responsible of these bands, we must compute the partial EDOS by summing the $|C_{\mu}^j(\vec{k})|^2$ for each element and each orbital. Here the $s$ and $p$~orbitals of germanium and sulfur atoms can be distinguished, as well as the $s$ orbitals of sodium atoms. We have scaled the partial EDOS so that that their sum is equal to the total EDOS. The results are illustrated in Fig.6, where the solid line represents the total EDOS and the dashed area shows the contribution of a given orbital.\\
It can be seen that zone A is almost exclusively caused by the $3s$ orbitals of sulfur atoms. The small peak at the end of this band, which appears to be shifted for x=0.5, has been attributed in GeS$_2$ to S-S homopolar bonds\cite{blaineau3}. We find that in the Na$_2$GeS$_3$ sample these homopolar bonds are also connected to a sodium atom, which changes thus the energy eigenvalues of these orbitals. This explains why the modes at the end of band A have a higher energy ($\approx$ 0.3 eV) for x=0.5 than in the GeS$_2$ compound. The band B, which is sharper for sodium-enriched systems (Fig.4), is mainly caused by the $4s$ orbitals of Ge atoms (here the respective concentrations of each element must be taken into account). This property has been observed experimentally, and has been attributed to the increase of the Ge-Ge interatomic length, limiting thus the height of band B \cite{foix}. We have analyzed the interatomic distances for Ge-Ge pairs, and we can confirm this experimental prediction. Indeed, the distance between germanium pairs increases in both edge-sharing and corner-sharing units. In GeS$_2$ these lengths were found to be equal respectively to 2.91 \AA~ and 3.41 \AA \cite{blaineau1}, whereas in the Na$_2$GeS$_3$ sample they become equal to 3.01 \AA~ and 3.67 \AA. Therefore this variation could indeed be responsible of the evolution of band B when the sodium concentration increases.\\
We see that the contribution of the sodium atoms in the EDOS appears at the beginning of band C. The electronic energy of the $3s$ orbitals of sodium atoms is therefore close to the energy of the $4p$ orbitals of germanium atoms, and both are responsible of the shoulder which can be distinguished at the beginning of this band. It can be seen at the end of band C that the last occupied states before the Fermi Level are caused by the $3p$ orbitals of sulfur atoms as in GeS$_2$, and no contribution of the Na ions could be determined at the top of the valence band.\\
The localization of the electronic eigenstates can be evaluated through the inverse participation ratio (IPR), which can be calculated with the Mulliken charges \cite{mulliken}. It appears that in comparison to the IPR of GeS$_2$, which has been shown in our previous work \cite{blaineau3}, no localized states concerning sodium atoms have been found. Indeed the IPR of Na$_2$GeS$_3$ is very similar to that of GeS$_2$ and therefore it is not shown here.\\
\section{Conclusion}
We have studied the vibrational and electronic properties of sodium thiogermanate glasses through DFT-based molecular dynamics simulations for different sodium concentrations. We find that the vibrational contribution of the sodium ions in the VDOS takes place between the acoustic and optic band. Thus, the modes in that zone are almost exclusively caused by the Na particles of the system, and are due to delocalized motions of the ions. In the same time, the acoustic band decreases, and in particular the well known ``soft modes'' tend to disappear because of the introduction of the sodium atoms in the system, which limits the collective vibrations connected to the Boson peak in that zone. This effect should be confirmed or infirmed experimentally. The calculation of the EDOS shows that the electronic contribution of Na in (x)Na$_2$S-(1-x)GeS$_2$ glasses is quite negligible. However, several variations could be seen in the three main contributions of the valence spectrum, and in particular in band B, which becomes sharper in the sodium-enriched systems. This change can be attributed to the increase of the Ge-Ge interatomic length, as it has been proposed experimentally. With the calculation of the partial EDOS we find that the contribution of the Na$^+$ ions takes place at the beginning of band C, in the same energy zone that the $p$ orbitals of the germanium atoms, and no localized states could be attributed to the sodium ions at the top of the valence band.\\
This work is the first step in the study of the influence of the alkali ions on the properties
of GeS$_2$ glasses and has to be continued by the analysis of the structural changes 
before trying to link all these modifications to the transport properties: this work is currently
in progress.\\
{\bf Acknowledgments} Parts of the simulations have been performed on the computers of
the ``Centre Informatique National de l'Enseignement Sup\'erieur'' (CINES) in Montpellier. 

\vskip 0,5cm
\begin{figure}[b]
\centerline{\includegraphics[width=11cm]{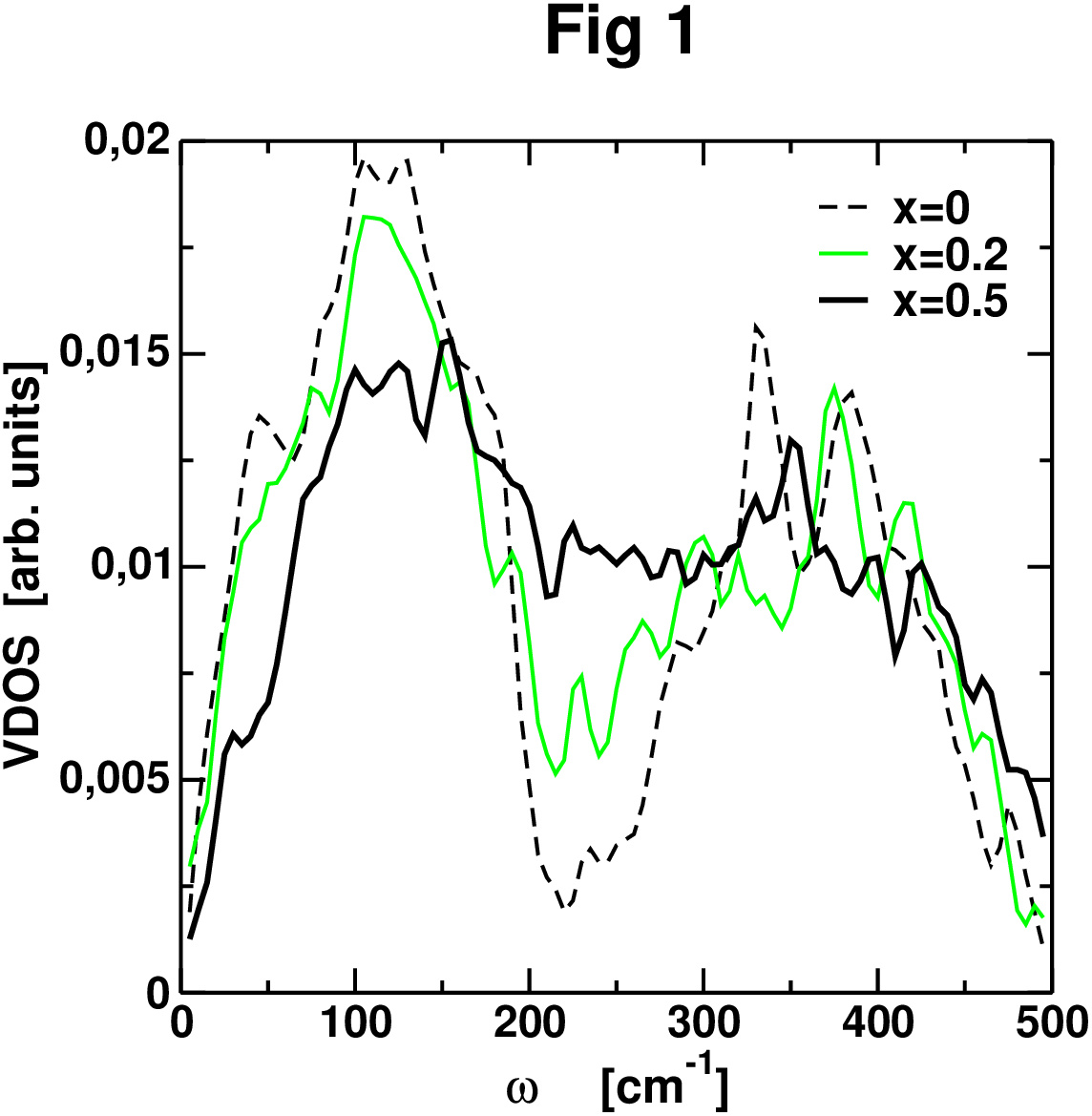}}
\caption{
Calculated VDOS of (x)Na$_2$S-(1-x)GeS$_2$ for x=0, x=0.2 and x=0.5.
}
\label{fig1}
\end{figure}

\newpage

\vspace*{2cm} 

\begin{figure}[h]
\centerline{\includegraphics[width=12cm]{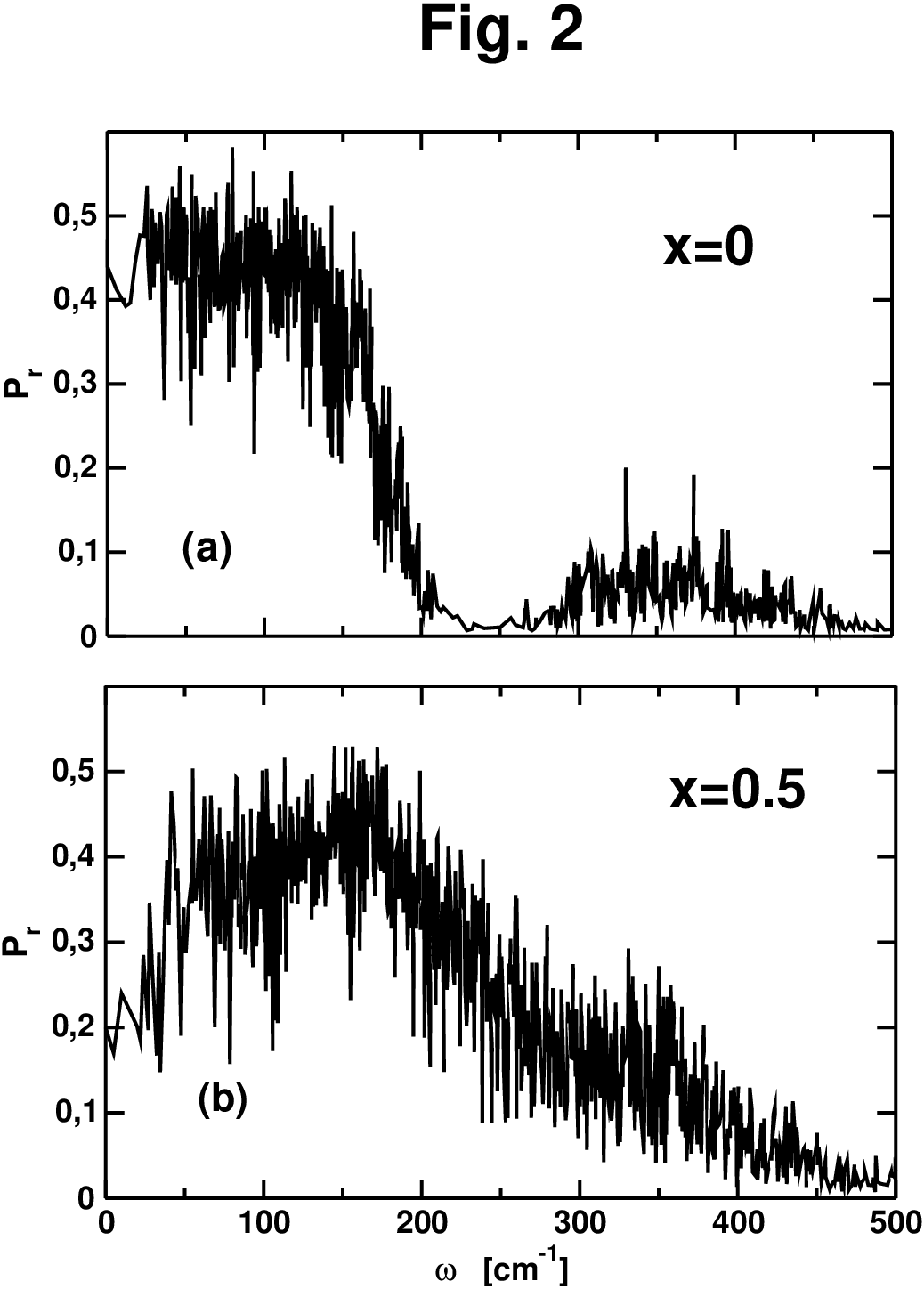}}
\caption{
Participation ratio (see text for definition) of $(a)$ GeS$_2$ (x=0), and $(b)$ Na$_2$GeS$_3$ (x=0.5), as a function of $\omega$.
}
\label{fig2}
\end{figure}

\newpage

\vspace*{2cm} 

\begin{figure}[h]
\centerline{\includegraphics[width=12cm]{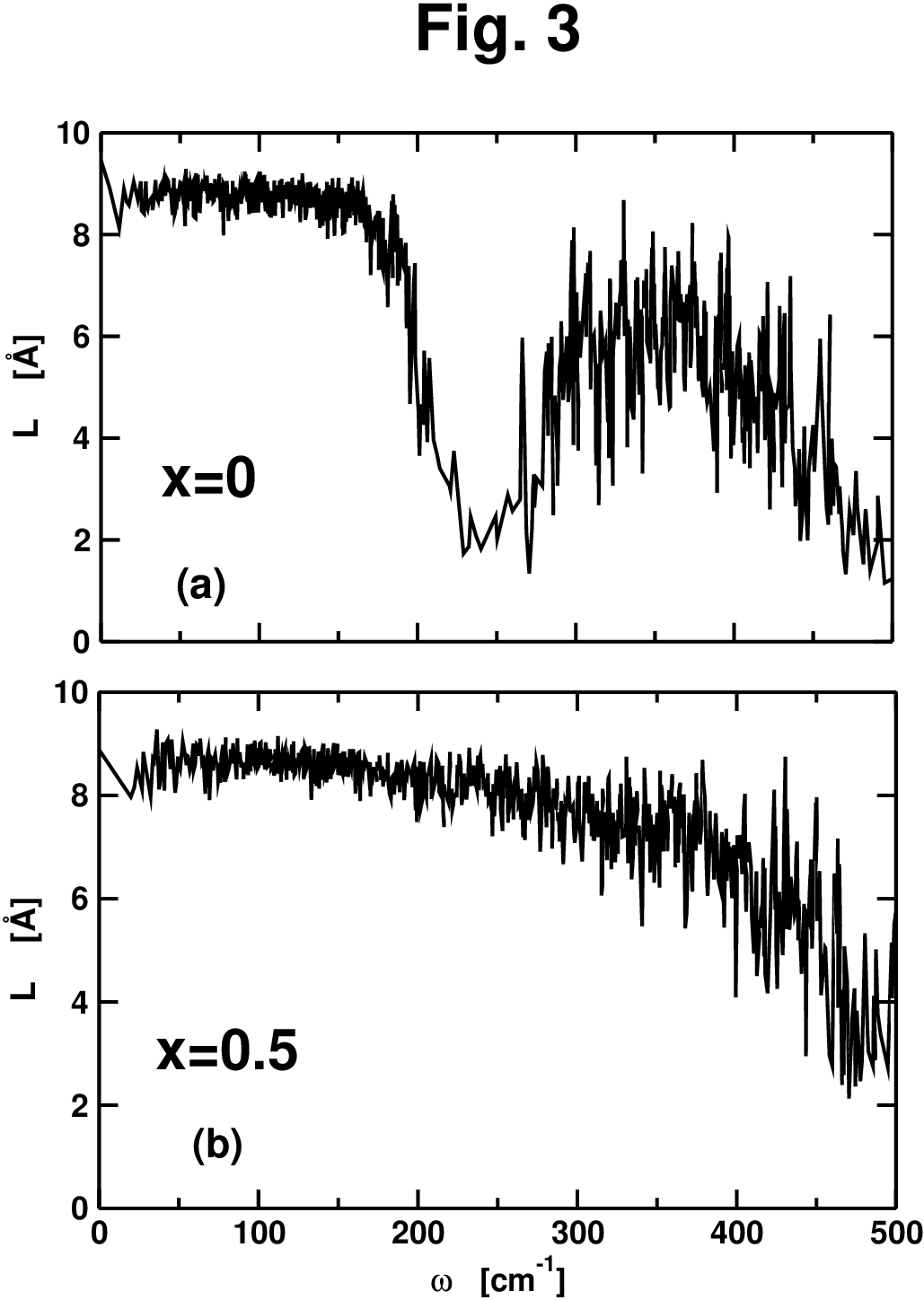}}
\caption{
Localization length (see text for definition) of GeS$_2$ (x=0) $(a)$, and Na$_2$GeS$_3$ (x=0.5) $(b)$, as a function of $\omega$.
}
\label{fig3}
\end{figure}
\newpage
\begin{figure}[t]
\centerline{\includegraphics[width=10cm]{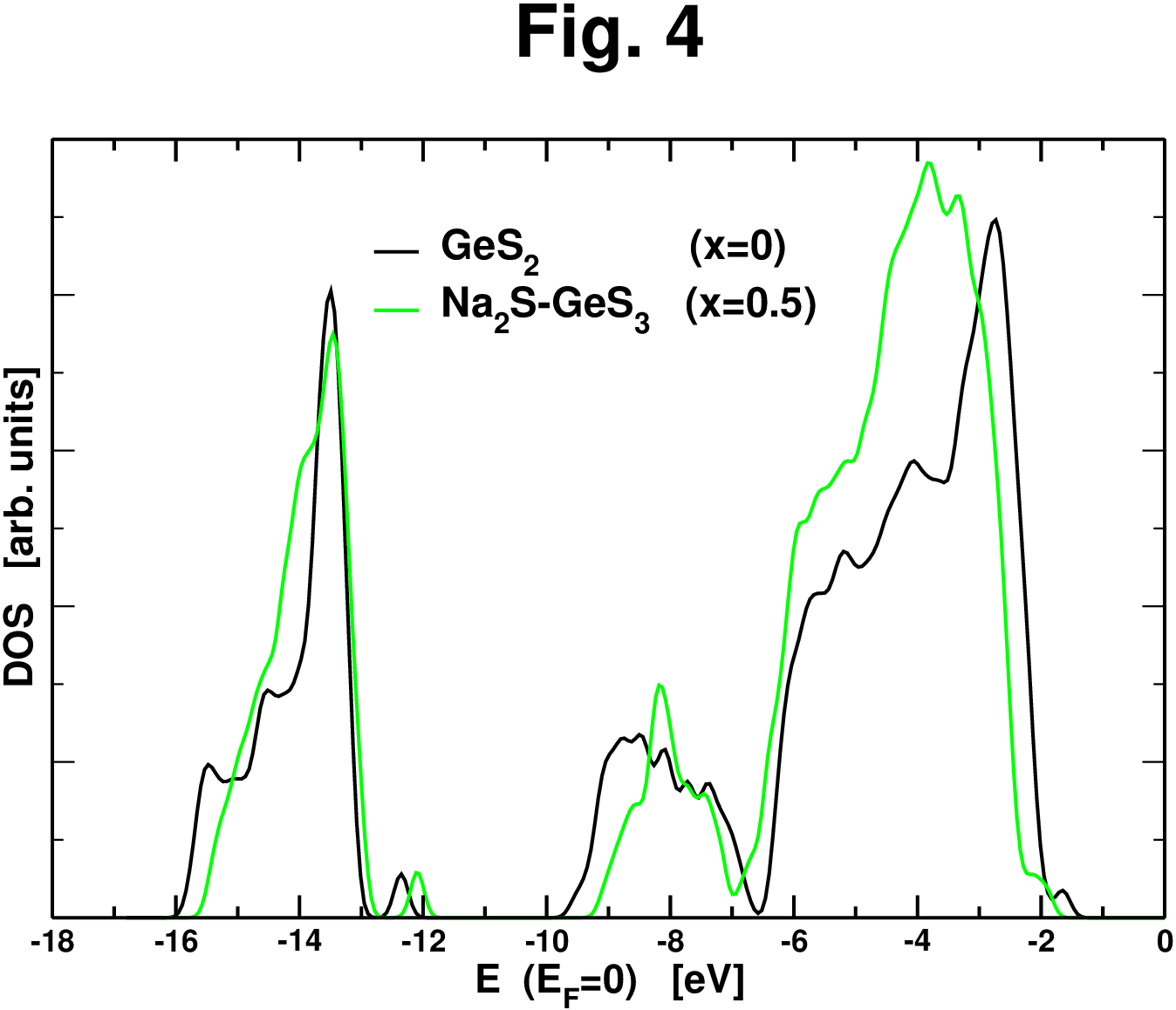}}
\caption{
Calculated EDOS of GeS$_2$ and Na$_2$GeS$_3$ $(a)$.
}
\label{fig4}
\end{figure}

\begin{figure}[b]
\centerline{\includegraphics[width=9cm]{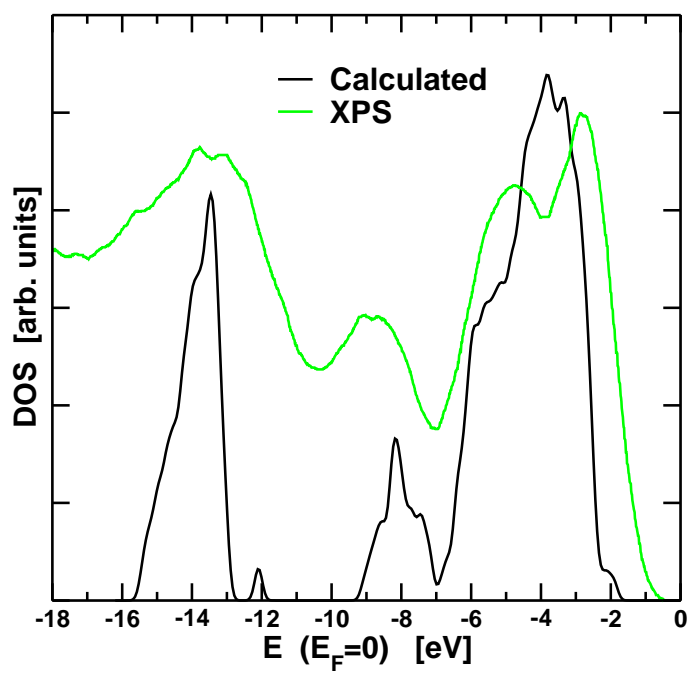}}
\caption{
Calculated EDOS of Na$_2$GeS$_3$ $(a)$ and experimental valence spectrum obtained by XPS measurements (Ref.18) .
}
\label{fig5}
\end{figure}

\newpage
\vspace*{1cm}

\begin{figure}[b]
\centerline{\includegraphics[width=8cm]{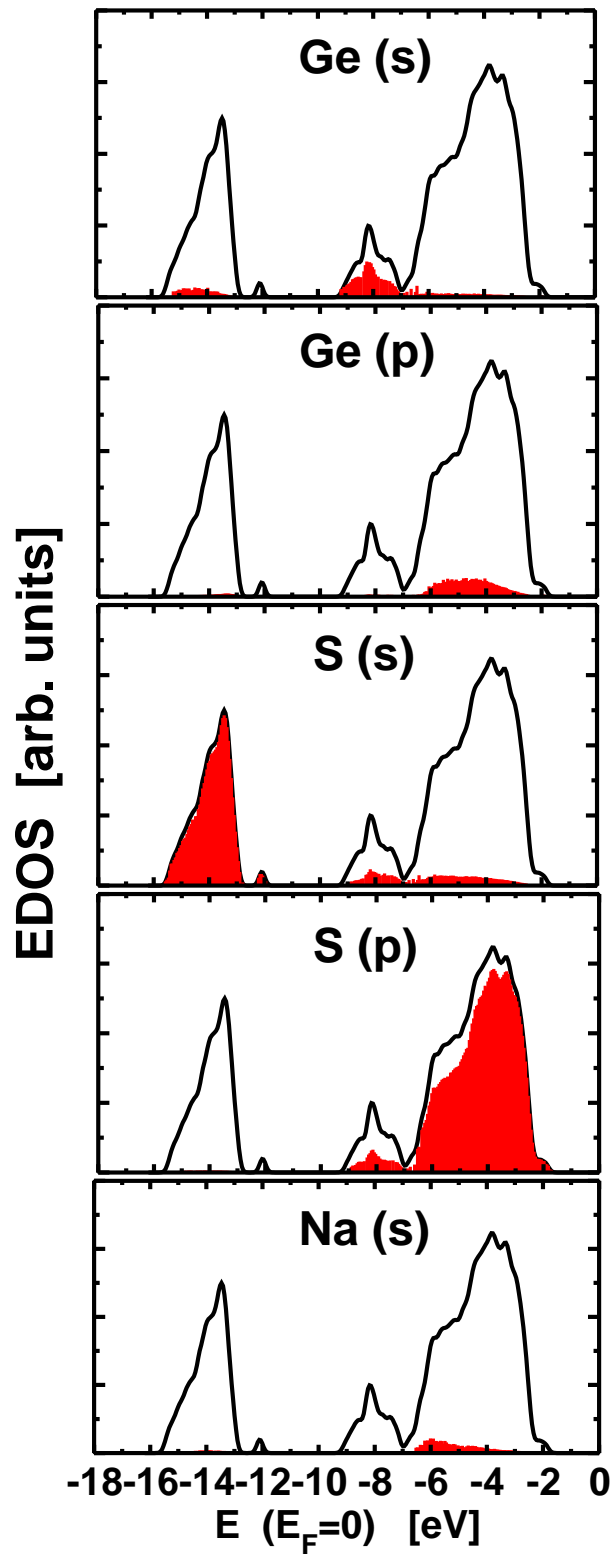}}
\caption{
Partial EDOS of Na$_2$GeS$_3$ (shaded area) and total EDOS (solid line).
}
\label{fig6}
\end{figure}
\hfill
\end{document}